\documentclass{aa}
\usepackage{graphicx}
\begin{document}

\title{Statistical characterisation of full-disk EUV/XUV solar irradiance and correlation with solar activity}

\author{J.~Greenhough\inst{1}\and S.~C.~Chapman\inst{1}\and R.~O.~Dendy\inst{2,1}\and V.~M.~Nakariakov\inst{1}\and G.~Rowlands\inst{1}}

\offprints{J.~Greenhough}

\institute{Department of Physics, University of Warwick, Coventry CV4 7AL, UK\\
\email{\{greenh,sandrac,valery\}@astro.warwick.ac.uk}
\email{g.rowlands@warwick.ac.uk}
\and UKAEA Culham Division, Culham Science Centre, Abingdon, Oxfordshire, OX14 3DB, UK\\
\email{richard.dendy@ukaea.org.uk}}

\date{Received 2003 May / Accepted 2003 }

\abstract{
We investigate the distribution of fluctuations in solar irradiance when integrated over the full disk, obtained using extreme ultraviolet/soft X-ray observations from the SOHO CELIAS/SEM instrument. This time series sums over both the contributions of single distinguishable flares, and of many other processes. By detrending we select events with timescales of less than a few hours such as waves, slow flows, and CMEs. The statistics generated by this range of phenomena can be characterised by power-law-tailed distributions. We show that (i) during the high activity period 2000 Jan--June the tail exponent $a_{T}=1.5\pm0.1$; (ii) during the low activity period 1996 Jan--June $a_{T}=3.0\pm0.2$; and (iii) in general $a_{T}$ decreases with increasing activity. 
}

\titlerunning{Full-disk EUV/XUV solar irradiance and solar activity}
\maketitle

\section{Introduction}\label{intro}
The statistical characteristics of solar irradiance have direct implications for coronal heating and terrestrial climate change; these statistics can be addressed both through flare events and, as in the present paper, through global measures. Many authors have identified transient flaring events and quantified their frequency distributions. The earliest observations of soft X-ray bursts by Drake (\cite{drake}), and of hard X-ray bursts by Datlowe et al. (\cite{datlowe}), Lin et al. (\cite{lin}), and Dennis (\cite{dennis}), found that the number of bursts with a given peak photon flux follows an inverse power law with logarithmic slope $a_{E}\approx 1.8$. These findings gave rise to the model of the active corona as the superposition of a large number of impulsive phenomena, from the largest \emph{flares} ($10^{33}$~ergs) down to \emph{nanoflares} ($10^{24}$~ergs) and possibly beyond. The contribution of nanoflares to coronal heating is reviewed by Berghmans \cite{berghmans}. Kucera et al. (\cite{kucera}) present evidence for changes in the upper limit of flare energies with the sizes of active regions, while secular variations are generally not seen. However, Bai (\cite{bai}) and Bromund et al. (\cite{bromund}) report a 154-day periodicity in hard X-ray burst distributions, and Wheatland (\cite{wheatland}) finds a variation in the flaring rate.   

In this paper we investigate a different statistic: the distribution of fluctuations in the irradiance from the full solar disk, using EUV/XUV observations by the SOHO CELIAS/SEM instrument (described in Sect.~\ref{data}). This global data encompasses both flaring and other activity such as waves, slow flows, and CMEs. Importantly, this also includes flare activity that could not be identified by event selection from images. By characterising the distributions with single power-law tails, we find that the exponent thereof differs by a factor of two between periods of high and low solar activity.

\section{SOHO/SEM Data}\label{data}
\begin{figure}
\resizebox{\hsize}{!}{\includegraphics{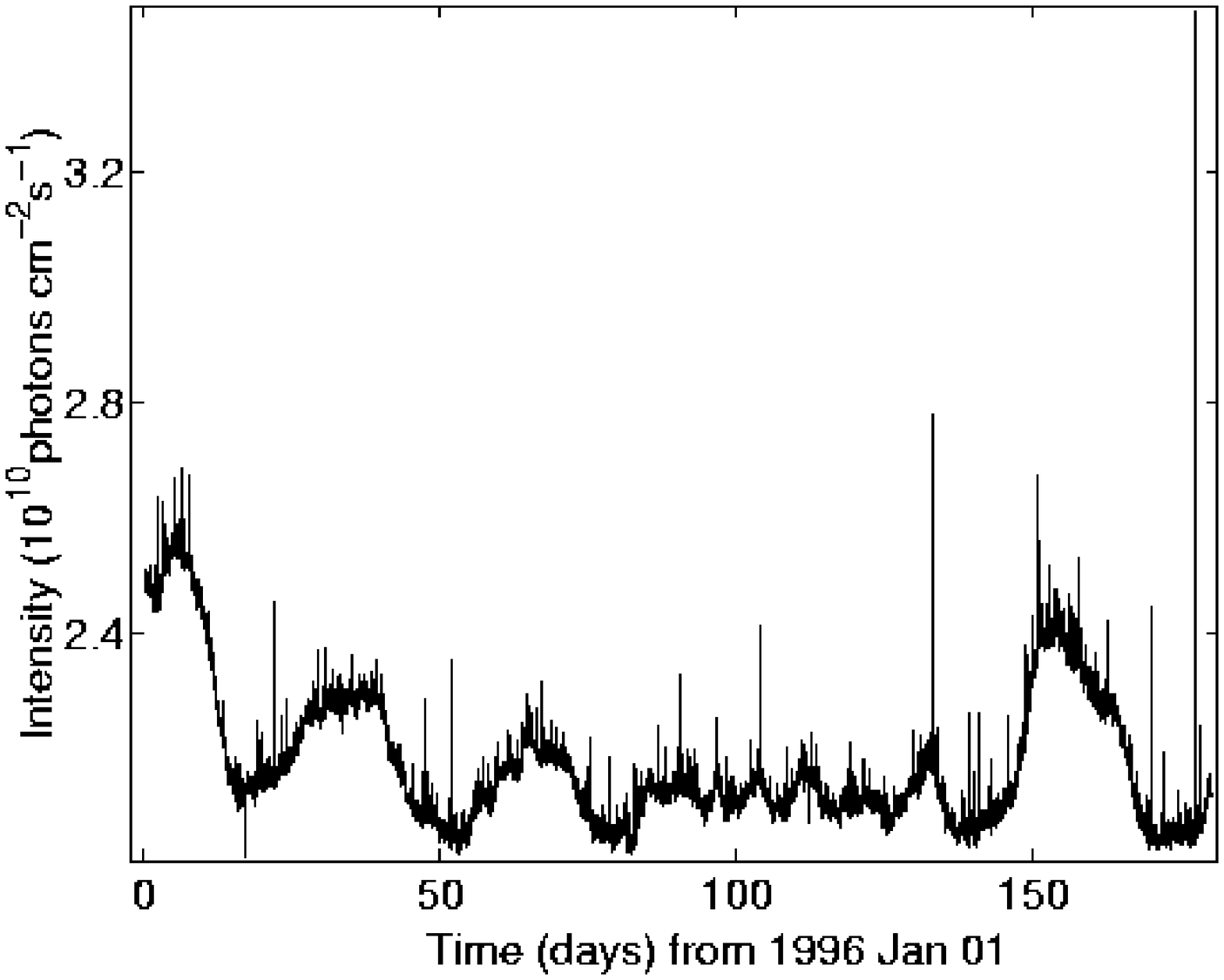}}
\resizebox{\hsize}{!}{\includegraphics{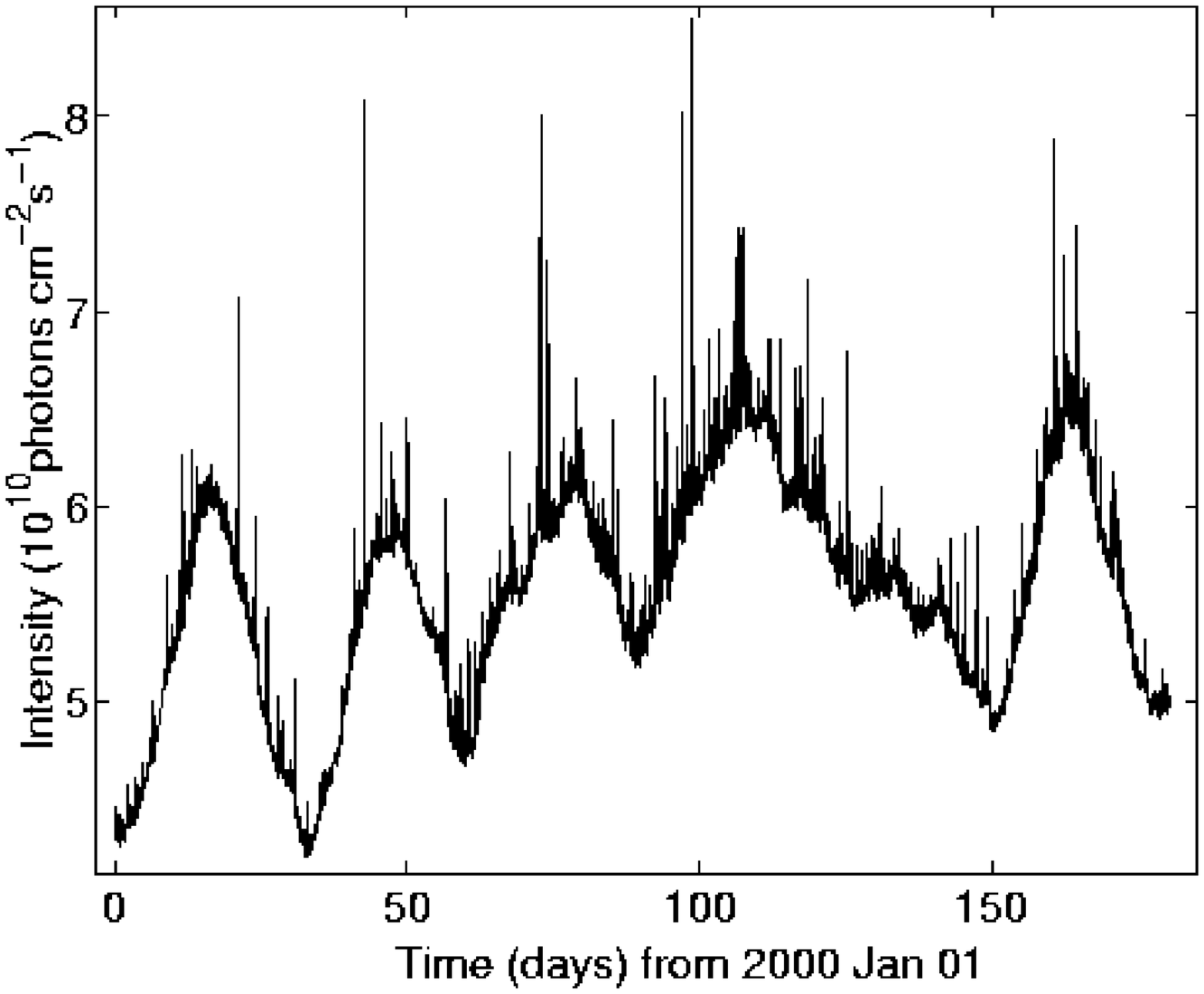}}
\caption{Time-series of raw full-disk solar flux (observed at 15~s intervals by SOHO/SEM); (upper) 1996 Jan--June (863450 measurements), (lower) 2000 Jan--June (1028342 measurements).\label{fig1}}
\end{figure}

The Charge, Element, and Isotope Analysis System/Solar Extreme Ultraviolet Monitor (CELIAS/SEM) transmission grating spectrometer, on board the Solar and Heliospheric Observatory (SOHO) spacecraft, has measured with 15~s resolution the full disk absolute photon flux at 1~AU since the launch of SOHO in 1995 Dec. Full details of the instrumentation and calibration are given by Hove et al. (\cite{hove}) and Judge et al. (\cite{judge1}), and the data are available via the internet\footnote{http://www.usc.edu/dept/space\_science/sem\_first.htm}. Previous EUV/XUV measurements suffered from instrument degradation, but the SEM data is highly stable and reliable with an estimated absolute uncertainty of $\sim\pm10\%$ ($1~\sigma$) (Judge et al. \cite{judge2}). Owing to the statistical correlation of electron temperature and thermal energy in flaring processes (Aschwanden \cite{asch4}), narrowband detectors provide biased estimates of flare energy distributions (Aschwanden \& Charbonneau \cite{asch1}); to overcome this problem, we choose the broadband (0.1--50~nm) channel 2 data. In Fig.~\ref{fig1} we plot the full-disk solar flux averaged over 15~s intervals from 1996 Jan--June (a period of low activity) and 2000 Jan--June (a period of high activity). 
The six or so quasiperiodic variations with the largest amplitude are caused by the movement of active regions across the disk with the sun's 27-day rotation period, plus contributions from the evolution of active regions over days to months (Fr\"{o}hlich \& Pap \cite{frohlich1}) and from the 11-year solar cycle (Willson \& Hudson \cite{willson}).

\begin{figure}
\resizebox{\hsize}{!}{\includegraphics{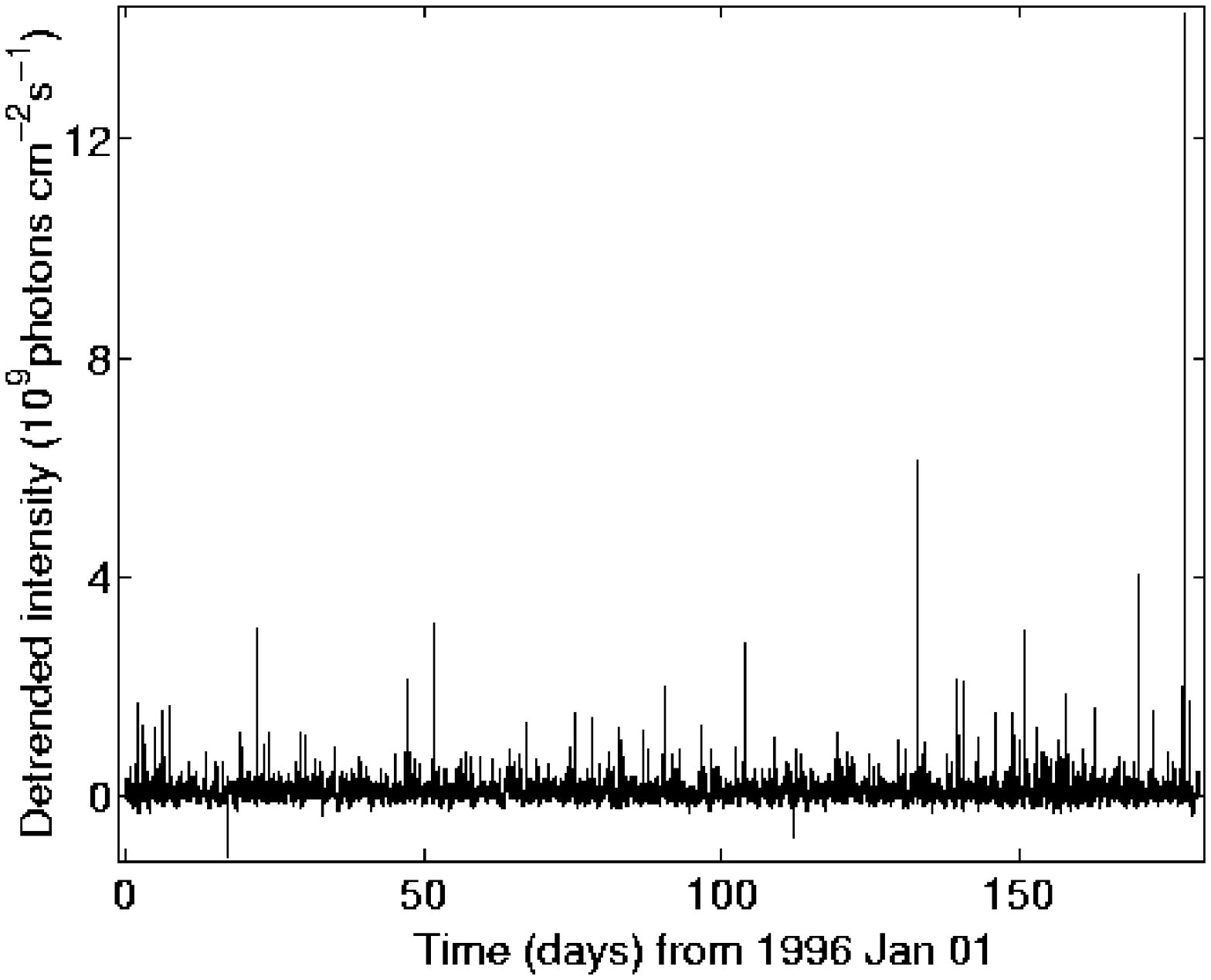}}
\resizebox{\hsize}{!}{\includegraphics{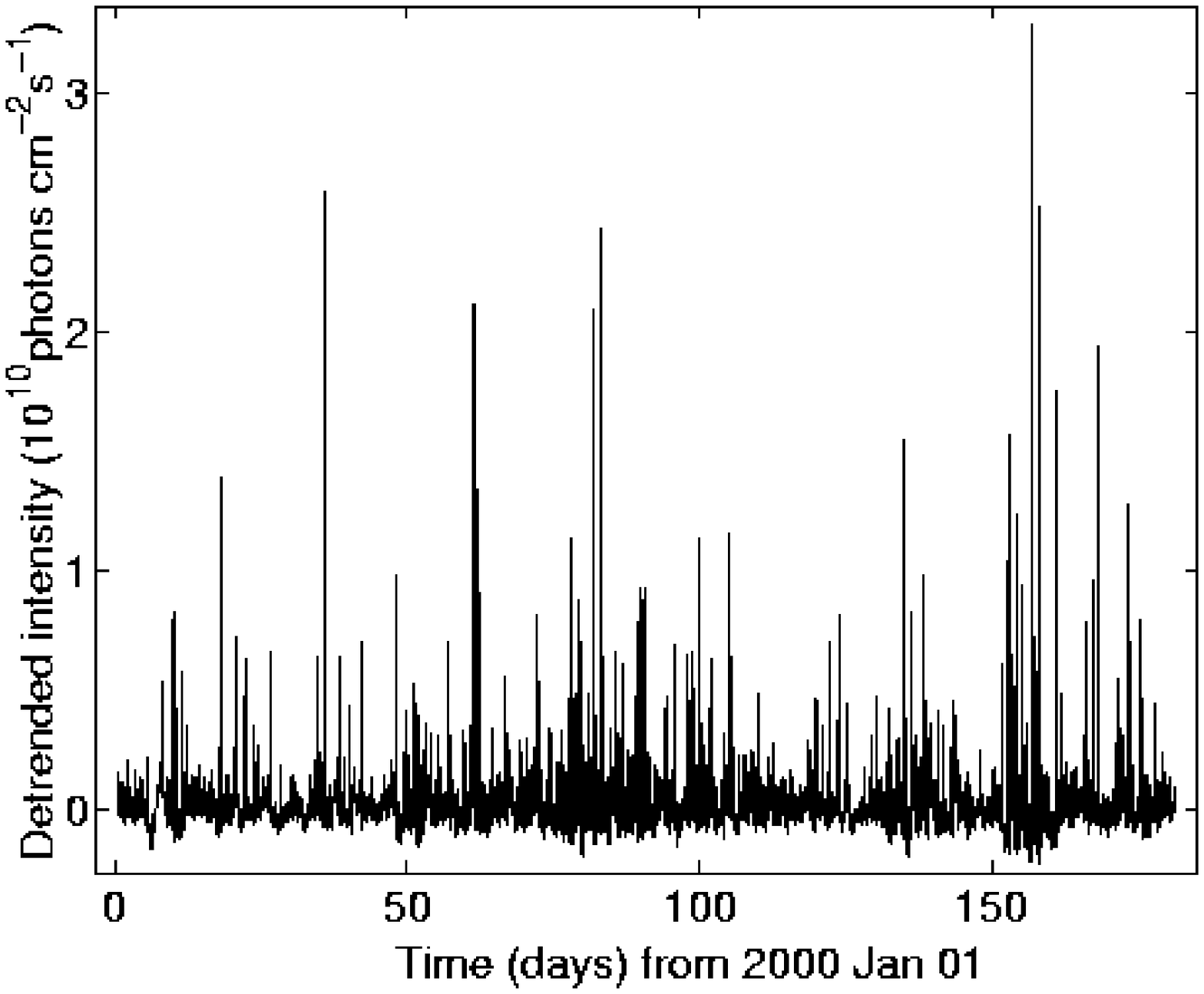}}
\caption{Time-series of raw full-disk solar flux (observed at 15~s intervals by SOHO/SEM) detrended by subtracting from each measurement the mean of the 10000 measurements (spanning $\tau=40$ hours) centred in time on that point; (upper) 1996 Jan--June, (lower) 2000 Jan--June.\label{fig2}}
\end{figure}

Since we wish to study only the short-time fluctuations, it is necessary to remove the long-term variations from the time-series. To achieve this we use a technique known as detrending, and the results are shown in Fig.~\ref{fig2}. 
Detrending is effected by subtracting from each data point the mean of the measurements over a certain time $\tau$ centred on that point. $\tau$ is chosen to be much less than the timescales to be removed and much greater than the timescales of interest. Here we set $\tau =40$ hours (10000 15~s sampling intervals) so as to remove the dominant 27-day cycle while retaining variations on timescales of up to a few hours.
 
\section{Full-disk irradiance distributions}\label{res}
\begin{figure}
\resizebox{\hsize}{!}{\includegraphics{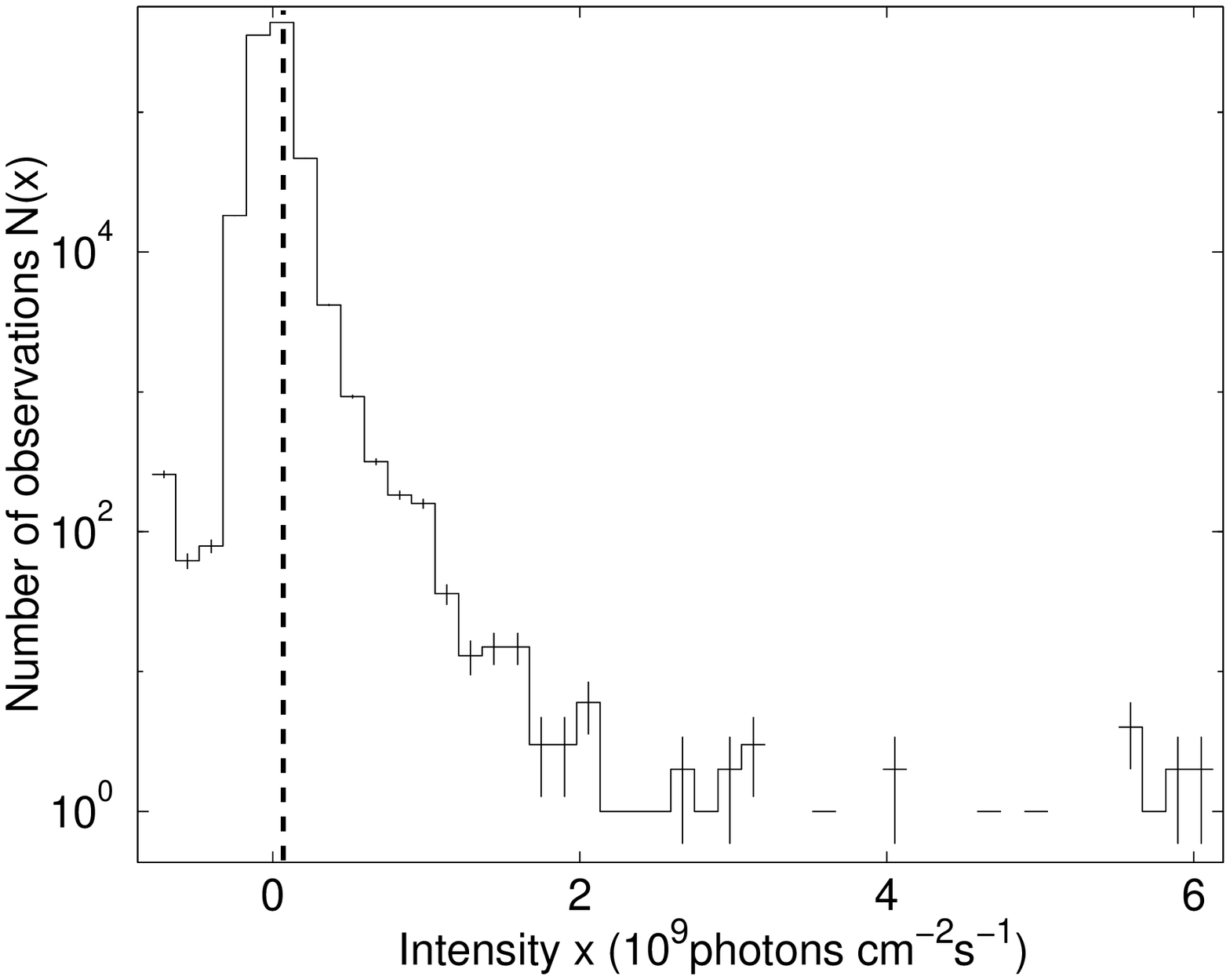}}
\resizebox{\hsize}{!}{\includegraphics{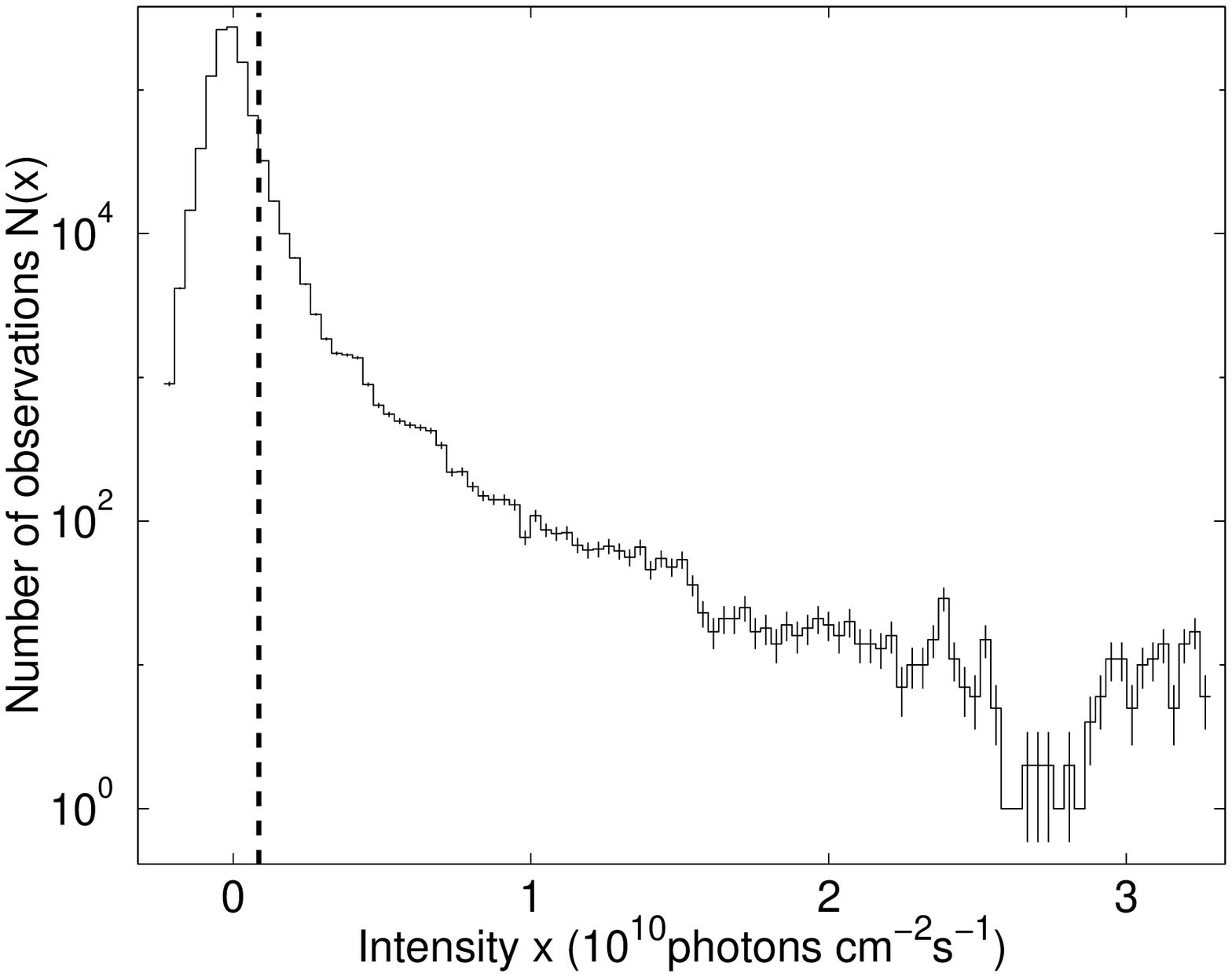}}
\caption{Log-linear distributions $\log N(x)$ of detrended full-disk solar flux $x$ (observed at 15~s intervals by SOHO/SEM); (upper) 1996 Jan--June, (lower) 2000 Jan--June. Vertical error bars indicate a factor of $\pm1/\sqrt{N(x)}$ for bins having $N(x)\ge 2$; bin widths (not shown for $N(x)=0$) indicate horizontal errors. Dashed lines indicate the lower limits for the fits in Fig.~\ref{fig4}.\label{fig3}}
\end{figure}

In Fig.~\ref{fig3} we plot the number distributions $N(x)$ of the detrended intensities $x$ during periods of low and high activity (shown as time-series in Fig.~\ref{fig2}), using log-linear axes to magnify the small probabilities in the tails of the distributions. 
It proved impossible to fit single distribution functions to the entire range of detrended data in this figure. Gaussian distributions, which are parabolic on these axes, clearly do not apply; log-normal distributions are inappropriate since many detrended values are negative. Fr\'{e}chet distributions (Chapman et al. \cite{chapman}; Greenhough et al. \cite{greenh}) may be fitted but they fit only $\sim 30\%$ of the bins within the errors shown in Fig.~3. Figure~4, which plots on log-log axes the ranges to the right of the dashed lines in Fig.~3, shows that the tails of the distributions approximate well to power laws. Single power laws of the form $N(x)\sim x^{-a_{T}}$, where $a_{T}$ is the tail exponent, can be fitted to give (i) $a_{T}=3.0\pm0.2$ for 1996 Jan--June, and (ii) $a_{T}=1.5\pm0.1$ for 2000 Jan--June, both having an $R^2$ goodness-of-fit of 0.99. 
\begin{figure}
\resizebox{\hsize}{!}{\includegraphics{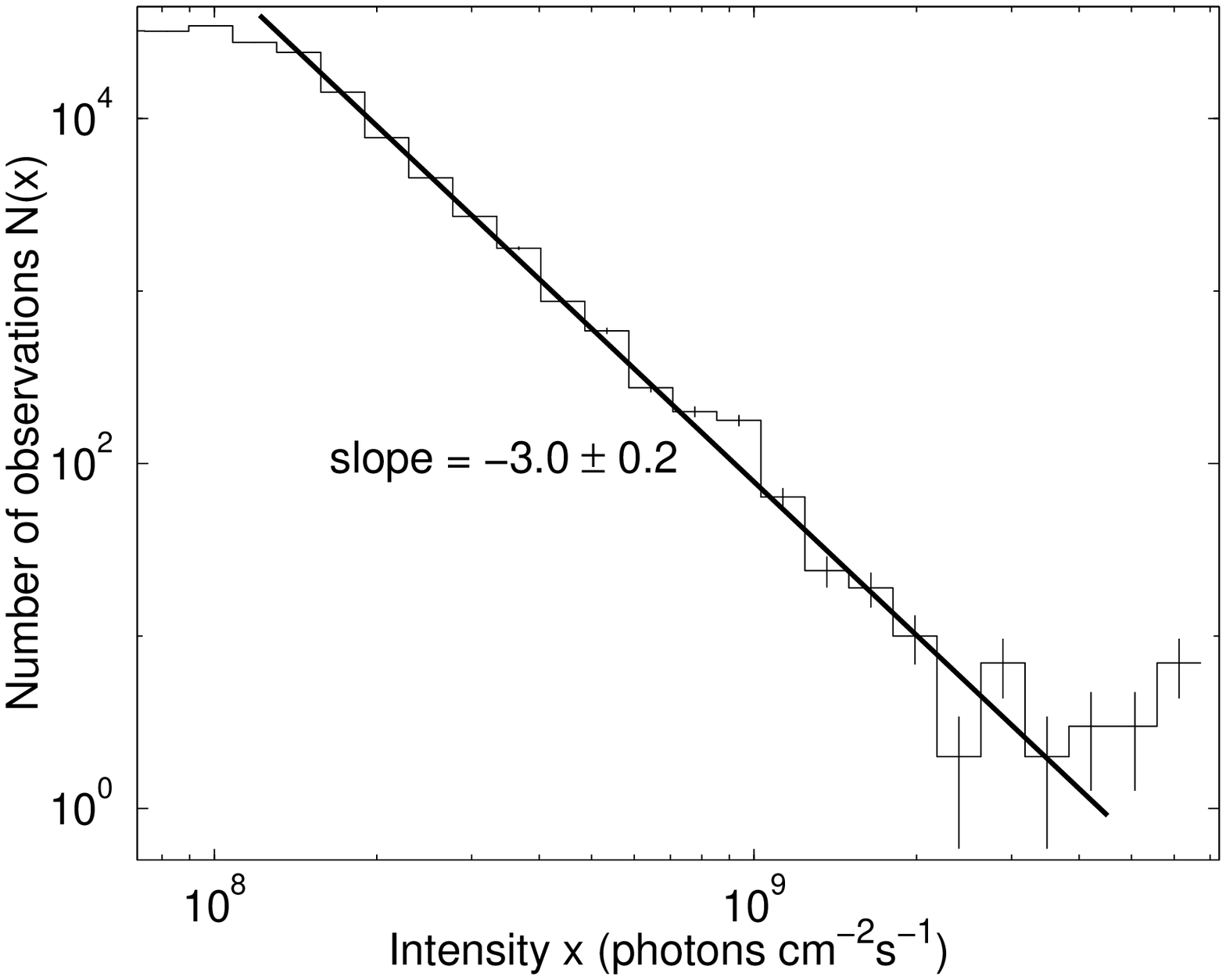}}
\resizebox{\hsize}{!}{\includegraphics{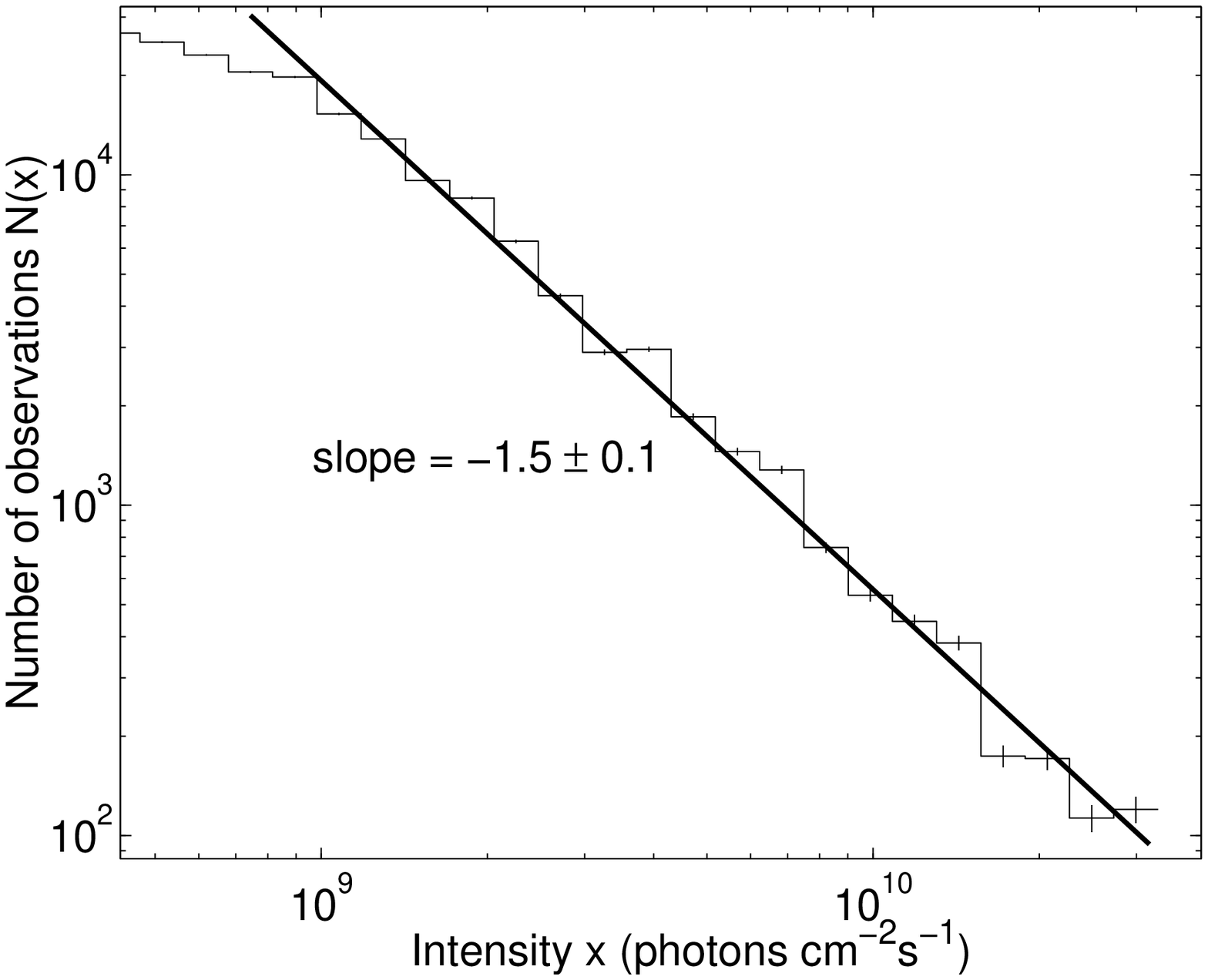}}
\caption{Log-log tails of distributions $\log N(x)$ of detrended full-disk solar flux $\log x$ (observed at 15~s intervals by SOHO/SEM); (upper) 1996 Jan--June, (lower)~2000 Jan--June. Vertical error bars indicate a factor of $\pm1/\sqrt{N(x)}$; bin widths indicate horizontal errors. Slopes $m$ are calculated via least-squares fitting with 95\% confidence intervals; tail exponent $a_{T}=-m$.\label{fig4}}
\end{figure}
To obtain the correct value for $a_{T}$ from the slope $m$ of $\log N(x)$ against $\log x$ (since $a_{T}=-m$), the data are placed into bins whose widths increase exponentially with $x$. One may also fit composite functions such as broken power laws. If the fitted slopes are allowed to change at, for example, $x=10^9$ (upper plot) and $x=5\times 10^9$ (lower plot), the best fits are respectively $a_{T}=2.8\pm0.2\rightarrow 3.3\pm1.2$ (overall $R^2=0.95$) and $a_{T}=1.3\pm0.2\rightarrow 1.5\pm0.2$ (overall $R^2=0.97$). Importantly, regardless of the details of the chosen fits, we find a significant difference in the tails of the detrended full-disk solar irradiance observations between the selected six-month periods of low and high activity. 
 
\begin{figure}
\resizebox{\hsize}{!}{\includegraphics{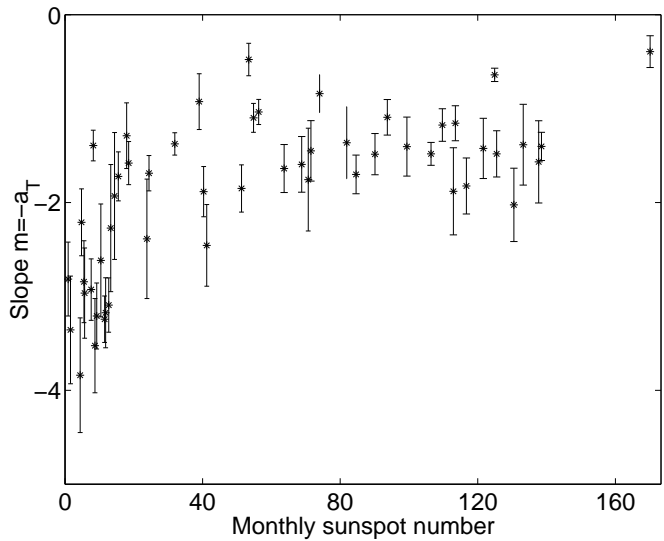}}
\caption{Slopes $m=-a_{T}$ of log-log distributions of detrended full-disk solar flux for each available month of data 1996 Jan--2000 Dec, plotted against the sunspot number for that month. Vertical error bars indicate 95\% confidence intervals for the least-squares fits used to obtain the values of $m$; three values whose errors span a greater range than the plot are omitted. \label{final}}
\end{figure}

In Fig.~\ref{final} we use single power laws as the simplest description of the distribution tails for intermediate times. We plot the slope $-a_{T}$ (with error bars showing $95\%$ least-squares confidence bands) for each available month of data from 1996 Jan to 2000 Dec, against the sunspot number for that month (a proxy for solar activity)\footnote{Available at http://sidc.oma.be/html/sunspot.html.}. 
Each value is calculated using a least-squares fit from the bin proceeding the mode up to the last non-empty bin in the log-log distribution of detrended data for that month. In three cases -- 1998 Oct, 1999 Feb, and 2000 Nov -- the error bars extend beyond the range of all the calculated slopes, and these points are not included in the figure. The plot is suggestive of a trend, albeit with a large variance, toward flatter power-law slopes as the sunspot number increases, which is consistent with the six-month averages presented in Fig.~\ref{fig4}. To our knowledge, these results represent the first clear activity-related distinction in the statistics of full-disk solar irradiation. 

\section{Conclusions and discussion}\label{discuss}
By removing the low-frequency (greater than a few hours) variations in full-disk EUV/XUV solar irradiance observations, and calculating the distributions of the remaining detrended fluctuations, we obtain distributions whose tails can be characterised by single power laws. The power-law exponent shows significant variation with activity: for the low-activity period 1996 Jan--June, we find $a_{T}=3.0\pm0.2$, while $a_{T}=1.5\pm0.1$ for the high-activity period 2000 Jan--June. Comparing $a_{T}$ and the sunspot number for each month of data suggests that there is in general a flattening of the slope with increasing activity. These results provide evidence for activity-related changes in irradiance fluctuations of the full solar disk. This is particularly intriguing given the diverse range of phenomena that contribute to this integrated emission measure. 

Changes in the value of $a_{T}$ may reflect important variations in the proportion or character of emissions from non-flaring regions. However, uncertainty is introduced by the dependence of the absolute photon flux on the relative spectral distribution, which may change during the intense flaring events (Ogawa et al. \cite{ogawa}) that are more frequent during high activity. One would have to remove the XUV component from the EUV/XUV total -- a process involving much uncertainty (Judge et al. \cite{judge2}) -- in order to test for this effect . Furthermore, since the relationship between total quantities and their underlying components is an unsolved problem in the field of non-equilibrium statistical mechanics (discussed by Chapman et al. (\cite{chapman}), and by Vekstein \& Jain (\cite{jain}) in the context of nanoflares), we do not attempt to make a quantitative link between $a_{T}$ and the flare statistic $a_{E}$ (see Sect.~\ref{intro}) in this paper.

\begin{acknowledgements}
We are grateful to Bogdan Hnat and Nick Watkins for helpful suggestions. J.~G. acknowledges a CASE Research Studentship from the UK Particle Physics and Astronomy Research Council in association with UKAEA. G.~R. acknowledges a Leverhulme Emeritus Fellowship. This work was also supported in part by the UK DTI. Data provided by the CELIAS/SEM experiment on the SOHO spacecraft, and the  Solar Influences Data analysis Center (SIDC, Royal Observatory of Belgium). 
\end{acknowledgements}

\end{document}